# Characterization of the Intrinsic and Extrinsic Resistances of a Microwave Graphene FET Under Zero Transconductance Conditions

Xiomara Ribero-Figueroa, Anibal Pacheco-Sanchez, Aida Mansouri, Pankaj Kumar, Omid Habibpour, Herbert Zirath, Roman Sordan, Francisco Pasadas, David Jiménez, and Reydezel Torres-Torres, *Senior Member, IEEE*

*Abstract*— Graphene field-effect transistors exhibit negligible transconductance under two scenarios: for any gate-to-source voltage when the drain-to-source voltage is set to zero, and for an arbitrary drain-to-source voltage provided that the gate-to-source voltage equals the Dirac voltage. Hence, extracting the channel and the parasitic series resistances from *S*-parameters under these conditions enables analyzing their dependence on the gate and drain biases. This is fundamental to assess the portion of the output resistance that is controlled by the gate. Besides, the drain bias dependence of the drain and source resistances is also evidenced. Within the proposal, resistive components accounting for the lossy nature of the gate capacitance are incorporated into the model, which exhibits a broadband correlation with experimental data. This avoids the series resistances to be considered as frequency dependent in the model.

## I. Introduction

GRAPHENE field-effect transistors (GFETs) are drawing attention in different applications [1]–[3]. Nevertheless, as for any other transistor, parasitic effects degrade their performance, such as the series resistances originated by the channel regions not electrostatically controlled by the gate electrode [4]–[7]. Hence, the circuit model of a GFET should consider the channel resistance ($R_{ch}$) associated with its intrinsic part, and also the resistances $R_g$, $R_s$, and $R_d$ related to the gate, source, and drain terminals, respectively. When GFET models are used for either device optimization or circuit design, accurately obtaining these resistances is mandatory [8]–[10].

A popular choice to analyze the resistive behavior of GFETs is through direct–current (DC) curves. However, DC methods are limited to characterize the total output resistance between the extrinsic terminals (i.e., $R_{out} = R_s + R_d + R_{ch}$). In this regard, regressions to data measured to transistors of several lengths are typically applied [11]–[12]. Unfortunately, the variance in the characteristics of different devices (e.g., in the contact resistance at the terminals) introduces errors in the parameter extraction [13]. Thus, a semi-analytical procedure was recently proposed to analyze $R_{out}$ from DC measurements of a single GFET [14], where gate-bias dependent and gate-bias independent parts are distinguished. Yet the whole bias-dependent contribution is included in $R_{ch}$, even though $R_s$ and $R_d$ may also exhibit a significant bias dependence [8],[15].

Alternatively, determining the resistances of FETs is possible from *S*-parameter measurements at zero-bias conditions (e.g., $V_{GS} = V_{DS} = 0$ V), by assuming that the device is turned off [16]. This technique, however, is not applicable to currently available GFETs because of the lack of bandgap in graphene [17]. Nonetheless, the resistive behavior of GFETs can be explored when the transconductance ($g_m$) is negligible [18]; this is the so-called cold-FET condition [19]. Conveniently, this 'zero-$g_m$' occurs in GFETs under two scenarios: 1) for $V_{DS} = 0$ V at any $V_{GS}$, and 2) when $V_{GS}$ equals the Dirac voltage ($V_{Dirac}$) for a given $V_{DS}$. The circuit elements of a GFET have been previously determined from RF measurements at $V_{DS} = 0$ V and $V_{GS} \neq 0$ V [20]; however, due to the large device's impedance at relatively low frequencies (LF), unexpected frequency dependence of $R_g$, $R_s$, and $R_d$ were reported. The latter points out the requirement of an improved model to achieve a meaningful parameter extraction, as presented in this paper.

Here, a methodology is proposed to obtain $R_g$, $R_s$, $R_d$, and $R_{ch}$ from *S*-parameters under zero-$g_m$ conditions, which allows to identify the dependence of these resistances on both, $V_{GS}$ and $V_{DS}$. Using these data, a model and parameter extraction to represent the gate-bias dependence of these resistances is developed. Besides, since $R_s$, and $R_d$ are also obtained at

Manuscript received March 21, 2023. This work was supported in part by Consejo Nacional de Humanidades, Ciencias y Tecnologías (CONAHCyT), Mexico under Grant 288875 and Grant 1014346, from European Union's Horizon 2020 research and innovation programme under grant agreement No GrapheneCore3 881603, from Ministerio de Ciencia, Innovación y Universidades under grant agreements RTI2018-097876-B-C21(MCIU/AEI/FEDER, UE), FJC2020-046213-I and PID2021-127840NB-I00 (MCIN/AEI/FEDER, UE) and from Junta de Andalucía – Consejería de Universidad, Investigación e Innovación - Project Energhene P21_00149 and PAIDI 2020 grant no. 20804. *Corresponding author no: X. Ribero-Figueroa*

Xiomara Ribero-Figueroa and Reydezel Torres-Torres are with INAOE, Puebla 72840, Mexico (e-mail: xiomis.93@gmail.com).

A. Pacheco-Sanchez and David Jiménez are with the Departament d'Enginyeria Electrònica, Escola d'Enginyeria, Universitat Autònoma de Barcelona, Bellaterra 08193, Spain.

A. Mansouri, H. Zirath and O. Habibpour are with the Microwave Electronics Laboratory, Department of Microtechnology and Nanoscience, Chalmers University of Technology, 41258 Gothenburg, Sweden.

P. Kumar and R. Sordan are with the L-NESS, Department of Physics, Politecnico di Milano, 22100 Como, Italy.

F. Pasadas is with the Pervasive Electronics Advanced Research Laboratory (PEARL), Departamento. de Electrónica y Tecnología de Computadores, Universidad de Granada, Granada, 18071, Spain.

Color versions of one or more of the figures in this paper are available online at http://ieeexplore.ieee.org.

Digital Object Identifier 10.1109/TED.2023.3311772

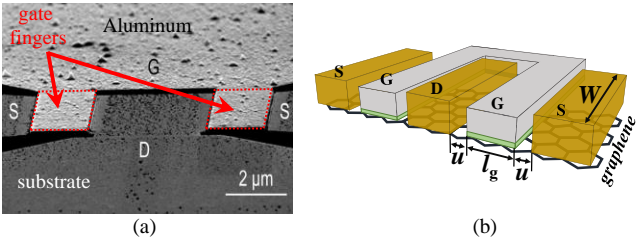
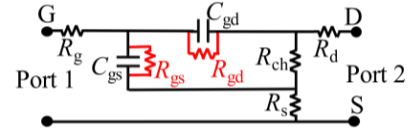

Fig. 1. Fabricated graphene transistor: a) SEM photograph, and b) simplified sketch not to scale for conceptually illustrating its structure, where gate length $l_g = 2$ µm, channel width $W = 10$ µm and $u = 20$ nm.

Fig. 2. Small-signal equivalent circuit for the GFET valid when the transconductance is negligible.

different $V_{DS}$ biases by applying the proposal, the significant dependence of these resistances on the drain bias is highlighted.

## II. DEVICE AND EXPERIMENTS

### A. Structure of the analyzed GFET

Fig. 1(a) shows a scanning electron microscope (SEM) image of the GFET, fabricated by electron-beam (e-beam) lithography, that was experimentally analyzed in this paper. For this device, the channel was made of monolayer graphene, which was grown by chemical vapor deposition on Cu, and then transferred to a high-resistivity Si substrate with a 1-µm-thick $SiO_2$ layer. The channels were defined by $O_2$ reactive-ion etching of wide-area graphene into 5-µm-wide graphene stripes. As depicted in Fig. 1 (b), the GFETs had two gate fingers of length $l_g = 2$ µm, obtained by evaporating 200 nm of Al in an e-beam evaporator at a base pressure of ~$10^{-6}$ mbar. Hence, the total transistor channel width is $W = 10$ µm. Besides, the gate insulator consists of a 4-nm-thick layer of $AlO_x$, which was formed at the interface between graphene and Al by exposing the samples to ambient air [21]. Finally, the source/drain contacts were obtained by evaporating 100 nm of Au at the same base pressure. In this case, no adhesion layer was used, which minimized the contact resistance [22].

The wideband characterization of GFETs of similar structure, as the one described here, is required for a wide range of applications. Additionally, accurately determining the corresponding series extrinsic parasitics is relevant, for instance, for resistive mixers [23]–[25] and zero-bias detectors [26]–[28]. Some of these applications have been demonstrated with 2D devices of micrometric gate lengths [26],[28]. In fact, one of the immediate niches for 2D devices are RF integrated circuits [29] where the device footprint is not a main concern.

### B. Measurements

The device is configured in common-source and is embedded between arrays of ground-signal-ground (GSG) pads. This allows for collecting the $S$-parameters up to 20 GHz using GSG RF probes with a pitch of 100 µm, which serve as interface to a vector network analyzer (VNA) setup. Regarding the biasing, two groups of measurements were performed; firstly, the GFET was biased at $V_{DS} = 0$ V and $V_{GS}$ was swept from –0.5 V to 1.5 V by steps of 0.25 V, whereas for the second group the $S$-parameters were collected at $V_{DS} = 0.3$ V and $V_{GS} = -0.25$ V, and then at $V_{DS} = 0.7$ V and $V_{GS} = 0$ V. These latter voltages were determined by fixing $V_{DS}$ and sweeping $V_{GS}$ to identify reciprocity complience (i.e., $S_{12} = S_{21}$), which implies that $V_{GS} = V_{Dirac}$ and the fulfillment of the zero-$g_m$ condition.

It is important to remark that the VNA setup was calibrated by applying an off-wafer line-reflect-match (LRM) algorithm to remove the effect of the cables and probes as well as for defining a reference impedance of 50 Ω. In addition, a two-step de-embedding procedure using the measurements of on-wafer "open" and "short" dummies was carried out to subtract the effect of the pad parasitics and other interconnects external to the transistor from the experimental data [30].

## III. ZERO-$G_M$ MODELING APPROACH

### A. Parameter extraction methodology

The circuit in Fig. 2 represents a GFET under the zero-$g_m$ conditions, where $C_{gs}$ and $C_{gd}$ are the gate-to-source and the gate-to-drain capacitances, respectively. On the other hand, the drain-to-source capacitance is neglected due to the long device length; this assumption is experimentally confirmed further below. Conversely, the resistors $R_{gs}$ and $R_{gd}$ are introduced here to account for the losses associated to $C_{gs}$ and $C_{gd}$.

To start with the parameter extraction, considering that the graphene channel exhibits low effective resistance when compared to that of the gate-to-channel insulator even at the Dirac voltage, $R_{ch} \ll R_{gs}$ and $R_{ch} \ll R_{gd}$ are assumed here. This yields that the effect of $R_{gs}$ and $R_{gd}$ is not significant on the transistor's output impedance. Hence, the imaginary part of the $Z_{22}$ parameter associated to the circuit in Fig. 2 is [18]:

$$\text{Im}(Z_{22}) = -\frac{\omega C_x R_{ch}^2}{1 + (R_{ch} C_x \omega)^2} \quad (1)$$

where $C_x = C_{gs} C_{gd}/(C_{gs} + C_{gd})$. In consequence, $R_{ch}$ and $C_x$ can be obtained at any bias condition where $g_m \approx 0$ through a linear regression of experimental data by involving (1) [18]. This enables the determination of the resistances $R_d$ and $R_s$ from [31]:

$$R_s \approx \text{Re}(Z_{21}) - 0.5A \quad (2)$$

$$R_d \approx \text{Re}(Z_{22}) - \text{Re}(Z_{21}) - 0.5A \quad (3)$$

where $A = -\text{Im}(Z_{22})/(R_{ch} C_x \omega)$. For the case of $R_g$, the extraction involves the $Z_{11}$ parameter, which is affected by the impact of $R_{gs}$ and $R_{gd}$. However, as for conventional FETs, at some gigahertz $1/(\omega C_{gs}) \ll R_{gs}$ and $1/(\omega C_{gd}) \ll R_{gd}$ can be assumed, implying that most of the alternating current flows through the capacitors; in this case, $R_g$ can be determined as:

$$R_g \approx [\text{Re}(Z_{11}) - \text{Re}(Z_{21})]_{HF} + 0.25A \quad (4)$$

where the 'HF' subscript indicates data at high frequencies. This frequency is about 10 GHz for the considered device.

After extracting $R_s$, $R_d$, and $R_{ch}$, for several $V_{GS}$ voltages, and fixed $V_{DS} = 0$ V, an expected gate-bias dependence is observed, which is analyzed in the next section. Moreover, since the proposed parameter extraction can also be applied at $V_{DS}$ voltages different from zero provided that $V_{GS} = V_{Dirac}$, a drain-bias dependence of these resistances can be identified. This fact is illustrated in Fig. 3, where $R_{ch}$ and the extracted extrinsic resistances are shown under the two possible scenarios of the zero-$g_m$ condition: at $V_{DS} = 0$ V, and at $V_{GS} = V_{Dirac}$. For comparison purposes, a fixed $V_{GS} = 0$ V is considered in these cases to allow attributing any change in the resistances to $V_{DS}$. Observe in this figure the sensibility of $R_s$, $R_d$, and $R_{ch}$ to $V_{DS}$, which is attributed to the nonlinear dependence on the drain-to-source voltage for the current flowing through the graphene. Notice the constant behavior over the frequency of the resistances, which unveils the adequacy of the parameter extraction. $R_g$ was determined at frequencies beyond 10 GHz, and no dependence on bias was observed since this resistance solely corresponds to the effect of the gate electrode [32]. Besides, its small magnitude is due to the wide gate electrode and the two-fingered structure.

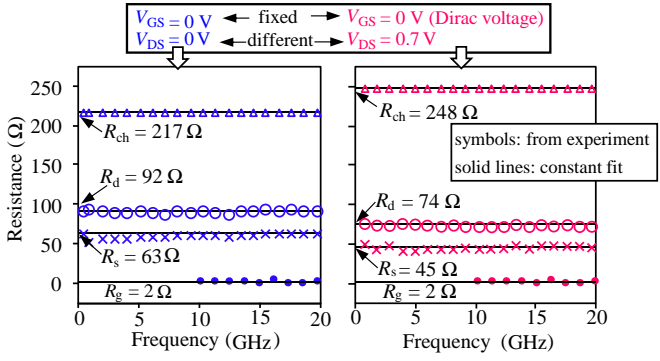

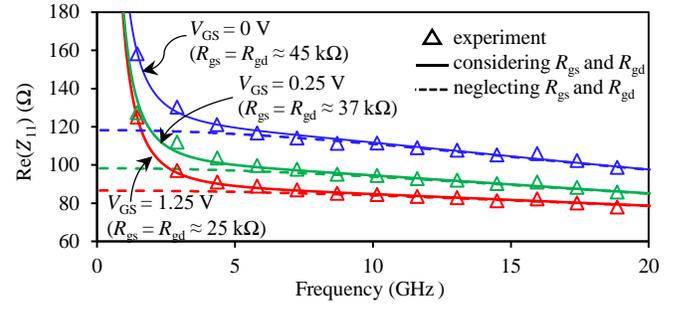

Fig. 3. Extraction of $R_g$, $R_s$, $R_d$, and $R_{ch}$ at $V_{GS} = 0$ V from the experimental Z-parameters at $V_{DS} = 0$ V (left), and at $V_{DS} = 0.7$ V (right).

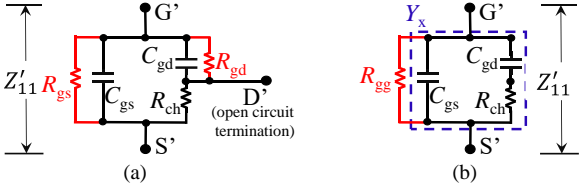

Fig. 4. Circuits representing $Z_{11}$ after removing the extrinsic resistances ($Z'_{11}$): a) original, b) re-arrangement obtained by assuming $R_{ch} \ll R_{gd}$.

$R_{gs}$ and $R_{gd}$ mainly impact the device's input port at relatively LF. Thus, the circuit in Fig. 4(a) can be deduced from the model in Fig. 2 by defining $Z'_{11}$ as the intrinsic parameter $Z_{11}$ (i.e., without the effect of $R_g$, $R_s$, and $R_d$). Hence, considering again that the conduction in the graphene channel makes the $R_{ch} \ll R_{gd}$ assumption to remain valid, the circuit for $Z'_{11}$ can be simplified to that in Fig. 4(b). From the later circuit, it can be inferred that $R_{gg}$ given by the parallel connection of $R_{gs}$ and $R_{gd}$ can be determined from experimental data at LF using:

$$R_{gg} = \text{Re}([1/Z_{11} - Y_x]_{LF})^{-1} \quad (5)$$

where $Y_x = j\omega(C_{gd}/(1+j\omega C_{gd}R_{ch}) + C_{gs})$ and data at frequencies below 5 GHz were considered for the device analyzed here. Moreover, as confirmed experimentally afterwards, the symmetry of the device when $g_m = 0$ yields $C_{gs} = C_{gd} = 0.5 C_x$ and $R_{gs} = R_{gd} = 2R_{gg}$.

Fig. 5. Re($Z_{11}$) versus frequency curves at $V_{DS} = 0$ V and different $V_{GS}$ illustrating the effect of including the resistive losses in the circuit of Fig. 2.

*B. Model–experiment correlation*

Firstly, to illustrate the requirement of considering $R_{gs}$ and $R_{gd}$ in the model shown in Fig. 2, the real part of the $Z_{11}$ parameter over frequency obtained with the model (cf. Fig. 2), using the extracted parameters, at different $V_{GS}$ are performed and compared with experimental data in Fig. 5. Notice that including these resistances improves the model–experiment correlation at frequencies below 10 GHz. Regarding this fact, $R_{gs}$ and $R_{gd}$ obtained here are as low as a few kiloohms, which points out that the loss associated to the gate dielectric is significant and observed at microwave frequencies; thus, the corresponding effect should be taken into account as in other FETs [33]. Besides, these resistances are further reduced as $V_{GS}$ increases since conduction mechanisms through the oxide are enhanced with the transverse electric field.

Now, before performing a full model–experiment correlation, the symmetry of the device without the effect of the extrinsic parasitics was verified by removing $R_g$, $R_d$, and $R_s$ from the experimental data, which allowed to obtain the intrinsic $Y'$-parameters. Afterwards, it was confirmed that $C_{gs} = \text{Im}(Y'_{11} + Y'_{12})/\omega$ yields approximately the same magnitude as $C_{gd} = -\text{Im}(Y'_{12})/\omega$. Additionally, it was observed that $C_{ds} = \text{Im}(Y'_{22} + Y'_{12})/\omega \approx 0$, which allows neglecting this capacitance in the circuit model.

At this point, the model including the intrinsic part and the series resistances can be implemented. Fig. 6(a) and Fig. 6(b) illustrate the Z-parameters for all the measured conditions, accurately reproduced by the model within the entire considered frequency range. In this regard, the two $g_m = 0$ conditions are covered: $V_{DS} = 0$ V for several $V_{GS}$, and $V_{DS} \neq 0$ V at $V_{GS} = V_{Dirac}$.

## IV. GATE BIAS-DEPENDENT MODELING

Whereas $R_g$ remains constant with bias as for FETs [34], the analysis of the dependence of $R_s$, $R_d$, and $R_{ch}$ on the gate bias is relevant for modeling a GFET. Thus, from the already known parameters at $V_{DS} = 0$ V, the gate-bias dependence of $R_s$, $R_d$, and $R_{ch}$, can be achieved in a simple manner. Bear in mind, however, that the ambipolar transfer characteristics of GFETs are different for $V_{GS} < V_{Dirac}$ (hole conduction) and $V_{GS} > V_{Dirac}$ (electron conduction). Hence, the model device parameters,

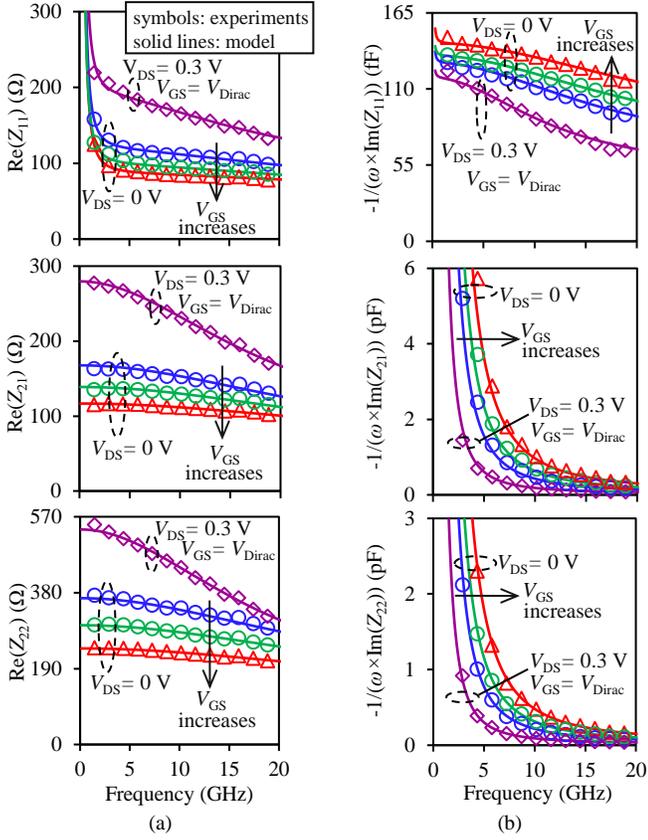

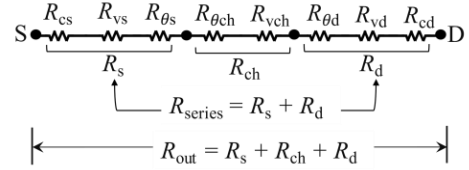

Fig. 7. Conceptual illustration of the different components for the resistances exhibited by the device between the drain and source terminals.

$$R_{V\gamma} = \frac{1}{\beta_\gamma \sqrt{V_{0\gamma}^2 + V_{GSO}^2}} \qquad (7)$$

where $\beta_\gamma$ is the transconductance parameter under low-field conditions, $V_{0\gamma}$ is related to the residual charge density, and

$$V_{GSO} = V_{GS} - V_{Dirac} - 0.5 V_{DS} \qquad (8)$$

$V_{Dirac}$ can obtained at the point where the output resistance versus $V_{GS}$ curve is at a maximum or, as carried out here, by identifying $V_{GS}$ at which $S_{12} = S_{21}$. Afterwards, the bias independent terms in (6) are grouped in $R_{CON\gamma} = R_{C\gamma} + R_{\theta\gamma}$, which combined with (7) allows for rewriting (6) as:

$$\frac{1}{(R_\gamma - R_{CON\gamma})^2} = (\beta_\gamma V_{0\gamma})^2 + \beta_\gamma^2 V_{GSO}^2 \qquad (9)$$

This equation represents a straight line with three unknowns: $R_{CON\gamma}$, $\beta$, and $V_0$. Therefore, performing a linear regression to $(R_\gamma - R_{CON\gamma})^{-2}$ versus $V_{GSO}^2$ data, $\beta_\gamma$ and $V_{0\gamma}$ are determined from calculations involving the corresponding intercept and slope. To perform this regression, $(R_\gamma - R_{CON\gamma})^{-2}$ versus $V_{GSO}^2$ data is firstly plotted by considering an arbitrarily small value for $R_{CON\gamma}$. Thus, when $R_{CON\gamma}$ is either underestimated or overestimated, the plotted data bend down or up, respectively, deviating from a linear trend. This is illustrated in Fig. 8 for $R_d$ for $V_{GS} \geq 0.25$ V, which allows assuming electron-only current. Fig. 8 shows that $R_{CONd}$ is modified until the coefficient of determination (referred to as $\rho^2$), associated with the linear regression, approaches unity.

## V. DISCUSSION

By using the procedure in Section IV, the gate-bias dependent model for $R_s$, $R_d$, and $R_{ch}$ was implemented. Afterwards, $R_{out}$ was obtained from the sum of these three components. Fig. 9 shows that agreement is achieved between this model and the data extracted from the measured S-parameters in the n-type region. Observe that the proposal allows for the identification of the separate contribution of each resistance component, evidencing the significant magnitude of the series resistances when compared to $R_{ch}$. Moreover, the separation of $R_s$ and $R_d$ components evidences a lack of perfect symmetry in the extrinsic elements due to differences in the metal-to-graphene contact on each side of the device. On the other hand, from the parameters listed in the inset of Fig. 9, the sum of the constant parts of $R_s$, $R_d$, and $R_{ch}$ yields 200 Ω, which explicitly indicates the part of $R_{out}$ not sensitive to $V_{GS}$. This resistance serves as a figure of merit that quantifies the portion of $R_{out}$ that only degrades the device's performance. Moreover, by using the such as the resistances, differ according to the region of interest [35]. For this reason, the model defined hereafter considers $V_{GS} \gg V_{Dirac}$, where dominant electron conduction occurs, but the methodology is equally valid for hole conduction at $V_{GS} \ll V_{Dirac}$. In fact, a combined analysis would also cover the transition region (i.e., $V_{GS}$ around $V_{Dirac}$) [14].

Fig. 6. Simulated Z-parameters confronted with experimental data. The cases correspond to $V_{DS} = 0$ V (with $V_{GS} = 0$ V, 0.25 V, and 1.25 V) and $V_{DS} = 0.3$ V (with $V_{GS} = V_{Dirac} = -0.25$ V): a) real parts, b) imaginary parts arranged as equivalent capacitances. Due to reciprocity when $g_m = 0$, $Z_{21}$ is omitted.

In a GFET, $R_{ch}$ is partly controlled by the gate voltage and also exhibits a bias independent component due to mobility degradation [8]. Furthermore, due to fringing gate capacitances, $R_s$ and $R_d$ comprise a gate-bias dependent part in addition to the constant contact and mobility degradation resistances. Thus, the total resistance between the drain and source terminals (i.e., $R_{out}$) can be represented using the circuit branch depicted in Fig. 7, where the components of $R_s$, $R_d$, and $R_{ch}$, are indicated. Mathematically, these resistances can thus be expressed as:

$$R_\gamma = R_{C\gamma} + R_{\theta\gamma} + R_{V\gamma} \qquad (6)$$

where $\gamma$ in the subscript is used to distinguish between the source (s), drain (d), and channel (ch) regions, and the three terms on the right-hand side of (6) are defined as follows. $R_{C\gamma}$ is the access resistance associated to the contact region and the ungated graphene region; thus, it is inferred that $R_{Cch} = 0$. The other resistance contribution independent of the gate voltage is $R_{\theta\gamma}$, which accounts for the mobility degradation within the graphene [14]. Finally, the bias-dependent term in (6) can be expressed as [11]:

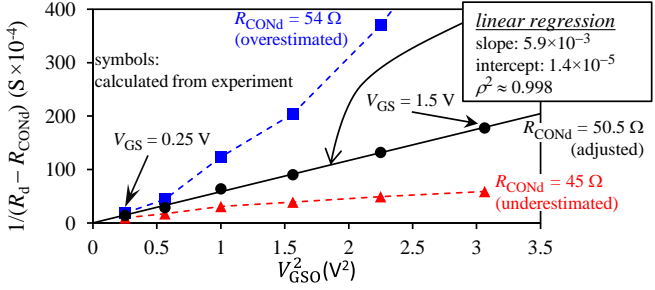

Fig. 8. Illustration of the parameter extraction for representing $R_d$ using the regression defined with the aid of (9).

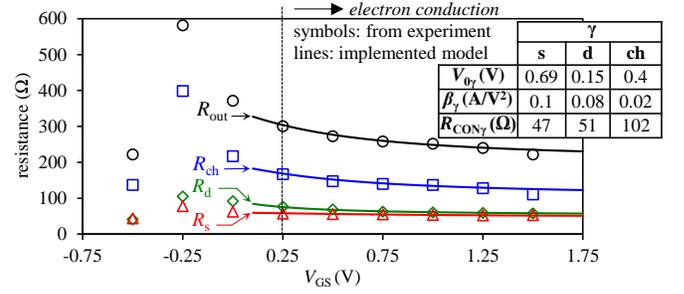

Fig. 9. Gate-bias dependent modeling the resistance components at $V_{DS} = 0$ V. The inset shows a table with the extracted model parameters.

proposal, it is now feasible realizing what part of the device is contributing more to increase this undesirable resistance.

The Fig. 10 presents the gate-voltage dependence of the $S$-parameters obtained employing the circuit model shown in Fig. 2 for two conditions where $g_m$ is negligible, and agreement with measurements is achieved. In fact, due to the reciprocity achieved under these conditions, only the data for $S_{21}$ are shown. For comparison purposes, curves assuming $R_s$, and $R_d$ to be independent of $V_{GS}$ and independent of $V_{DS}$ in Fig. 10 are included. The noticeable discrepancy of these later curves from experimental data points out the relevance of determining the extrinsic resistances at the desired bias conditions. Finally, for reference, the Table I was added to show the determined values for the model elements in Fig. 2, for different bias conditions.

## VI. CONCLUSION

The zero-$g_m$ parameter extraction allows for analyzing the gate and drain bias dependence of the gate, channel, and series resistances from $S$-parameters. During the analysis, it was observed that solely considering the gate, channel, and source resistances allow for accurately modeling the GFET's input impedance at frequencies above a few gigahertzes. Conversely, the operation at relatively low frequencies requires taking into consideration the loss associated to the gate capacitance. The latter consideration improves the description of the GFET small-signal response over traditional approaches, which neglect such phenomena. Furthermore, by implementing a gate-bias dependent model considering mobility degradation in the graphene, the constant and gate-bias dependent parts of the source, drain, and channel resistances are quantified and represented, which allows reproducing broadband $S$-parameters. This provides relevant information to device engineers to identify the characteristics to be improved when optimizing the structure and material properties of given GFET.

The data extracted using the proposal can be used to ease the implementation of models under conditions where the transconductance effect is not negligible. This is due to the establishment of reasonable limits during parameter determinations through curve optimization. Furthermore, the approach can be exploited for the technology evaluation and design of high-frequency applications where GFETs operate under passive conditions, e.g., in resistive mixers, for RF power harvesting, or where the GFET is biased at the Dirac voltage, such as in subharmonic mixers and frequency multipliers.

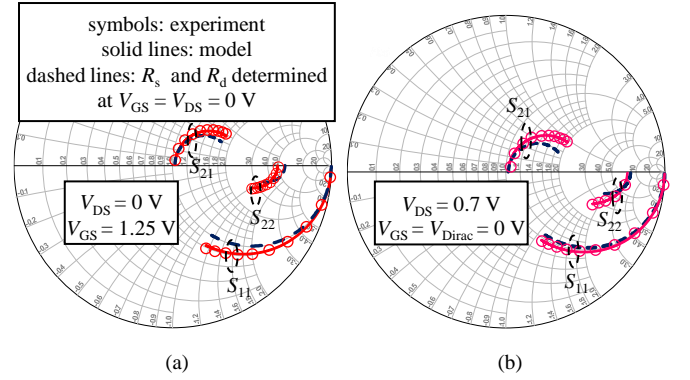

Fig. 10. Model–experiment correlation the $S$-parameters: a) at $V_{DS} = 0$ V and $V_{GS} = 1.25$ V (i.e., well above the Dirac voltage), and b) at $V_{GS} = 0$ V and $V_{DS} = 0.7$ V (i.e., at the Dirac voltage for this $V_{DS}$). In both plots, the dashed curves illustrate the model deviation from the experimental data when assuming the series resistances as bias independent obtained at $V_{DS} = V_{GS} = 0$ V.

TABLE I
EXTRACTED VALUES OF SMALL-SIGNAL CIRCUIT SHOWN IN FIG. 2.

| $V_{DS}$ (V) | $V_{GS}$ (V) | $R_g$ (Ω) | $R_s$ (Ω) | $R_d$ (Ω) | $R_{ch}$ (Ω) | $C_{gs}$ (fF) | $R_{gs}$ (kΩ) |
|---|---|---|---|---|---|---|---|
| **0** | **0** | 2 | 63 | 92 | 217 | 60 | 45 |
| **0** | **0.25** | 2 | 57 | 77 | 167 | 67.6 | 37 |
| **0** | **1.25** | 2 | 52.5 | 59.2 | 129 | 70 | 25 |
| **0.3** | **-0.25** | 2 | 69.5 | 47.5 | 422 | 40.6 | 79 |
| **0.7** | **0.0** | 2 | 74 | 45 | 248 | 60 | 51 |

## REFERENCES

[1] N. Yogeswaran, S. Gupta, and R. Dahiya, "Low voltage graphene FET based pressure sensor," in *2018 IEEE SENSORS*, New Delhi, India, Oct. 2018, pp. 1–4, doi: 10.1109/ICSENS.2018.8589569.

[2] M. Hajizadegan, M. Sakhdari, S. Abbasi, and P.-Y. Chen, "Machine learning assisted multi-functional graphene-based harmonic sensors," *IEEE Sensors J.*, vol. 21, no. 6, pp. 8333–8340, Mar. 2021, doi: 10.1109/JSEN.2020.3046455.

[3] M. Saeed, P. Palacios, M.-D. We, E. Baskent, C.-Y. Fan, B. Uzlu, K.-T. Wang, A. Hemmetter, Z. Wang, D. Neumaier, M. C. Lemme, R. Negra, "Graphene-based microwave circuits: a review", *Advanced Materials*, vol. 34, no. 48, Art. no. 2108473, Dec. 2022, doi:10.1002/adma.202108473.

[4] F. Zárate-Rincón, G. A. Álvarez-Botero, R. S. Murphy-Arteaga, R. Torres-Torres and A. Ortiz-Conde, "Impact of multi-finger geometry on the extrinsic parasitic resistances of microwave MOSFETs," in *IEEE MTT-S Int. Microw. Symp. Dig.*, Tampa,
© 2023 IEEE. Personal use of this material is permitted. Permission from IEEE must be obtained for all other uses, in any current or future media, including reprinting/republishing this material for advertising or promotional purposes, creating new collective works, for resale or redistribution to servers or lists, or reuse of any copyrighted component of this work in other works. DOI: 10.1109/TED.2023.3311772